\documentclass[12pt]{article}
\pdfoutput=1
\usepackage[utf8]{inputenc}
\usepackage{amsfonts,amsmath}
\usepackage{jheppub}

\usepackage{color}
\usepackage{tabularx}

\usepackage{ifthen}
\usepackage{enumerate}
\usepackage{amsmath}
\usepackage{amssymb}
\usepackage[mathscr]{eucal}
\usepackage[errorshow]{tracefnt}
\usepackage{amsmath}
\usepackage{slashed}
\usepackage{amssymb}
\usepackage{array}
\usepackage{amsthm}
\usepackage{eucal}
\usepackage{epsf}
\usepackage{epsfig}
\usepackage{psfrag}
\usepackage{multirow}
\usepackage{wasysym} 
\usepackage{tikz-cd} 
\usepgfmodule{shapes}
\usetikzlibrary{backgrounds, arrows,calc,shapes,decorations.pathreplacing, decorations.markings, automata,positioning}

\newcommand{\beqa}{\begin{eqnarray}}
\newcommand{\eeqa}{\end{eqnarray}}

\newcommand{\beq}{\begin{equation}}
\newcommand{\eeq}{\end{equation}}
\newcommand{\bea}{\begin{equation}\begin{aligned}}
\newcommand{\eea}{\end{aligned}\end{equation}}
\newcommand{\ba}{\begin{array}}
\newcommand{\ea}{\end{array}}

\newcommand{\M}{{\mathfrak M}}
\newcommand{\CN}{{\mathcal N}}

\newcommand{\CT}{{\mathcal T}}

\newcommand{\CW}{{\mathcal W}}
\newcommand{\CZ}{{\mathcal Z}}

\newcommand\Qt{\widetilde Q}
\newcommand\Pt{\widetilde P}
\newcommand\Rt{\widetilde R}

\newcommand\Qd{Q^{\dagger}}

\newcommand\Md{M^{\dagger}}

\newcommand{\be}{\begin{equation}}
\newcommand{\ee}{\end{equation}}

\newcommand{\bpic}{\begin{tikzpicture}}
\newcommand{\epic}{\end{tikzpicture}}

\def\+{{+\!\!\!+}}

\def\a{\alpha} 
\def\b{\beta}

\def\0{\nonumber}

\def\1{{\bf 1}}

\newcommand{\cC}{\mathcal{C}}

\newcommand{\cN}{\mathcal{N}}

\newcommand{\cW}{\mathcal{W}}

\newcommand{\bZ}{\mathbb{Z}}

\newcommand{\so}{\mathfrak{so}}

\newcommand{\ie}{\textit{i.e.}}
\newcommand{\matht}[1]{\ensuremath{\boldsymbol{#1}}}
\newcommand{\rep}[1]{{\bf #1}}
\DeclareMathOperator{\re}{\mathbb{R}e}
\DeclareMathOperator{\im}{\mathbb{I}m}

\newcommand{\sutf}{SU(2)_{\text{flav}}}
\newcommand{\uot}{U(1)_{\text{top}}}

\title{\matht{\CN{=}1} QED in \matht{2{+}1} dimensions: \\ Dualities and enhanced symmetries}

\preprint{SISSA /2018/FISI}

\author{Francesco Benini, }
\author{Sergio Benvenuti}

\affiliation{SISSA, Via Bonomea 265, 34136 Trieste, Italy}
\affiliation{INFN, Sezione di Trieste, Via Valerio 2, 34127 Trieste, Italy}
\affiliation{ICTP, Strada Costiera 11, 34151 Trieste, Italy}
\emailAdd{fbenini@sissa.it, benve79@gmail.com}

\abstract{We consider three-dimensional sQED with 2 flavors and minimal supersymmetry. We discuss various models which are dual to Gross-Neveu-Yukawa theories. The $U(2)$ ultraviolet global symmetry is often enhanced in the infrared, for instance to $O(4)$ or $SU(3)$. This is analogous to the conjectured behaviour of non-supersymmetric QED with 2 flavors. A perturbative analysis of the Gross-Neveu-Yukawa models in the $D = 4 - \varepsilon$ expansion shows that the $U(2)$ preserving superpotential deformations of the sQED (modulo tuning mass terms to zero) are irrelevant, so the fixed points with enhanced symmetry are stable. We also construct an example of $\mathcal{N} = 2$ sQED with 4 flavors that exhibits enhanced $SO(6)$ symmetry.}

\begin{document}

\maketitle

\section{Introduction and results}

An interesting feature of gauge theories in $2+1$ dimensions is infrared (IR) dualities that exchange standard mesonic operators with monopole operators. The first example was found long time ago: the bosonic particle-vortex duality \cite{Peskin:1977kp, Dasgupta:1981zz} states that a $U(1)$ gauge theory (QED) with one scalar flavor is dual to a theory without gauge fields: the $O(2)$ vector model.

In this paper we study similar examples, in the case of QED with minimal $\cN=1$ supersymmetry (SUSY). Recently, new dualities for $\CN=1$ non-Abelian gauge theories have been investigated \cite{Bashmakov:2018wts,Benini:2018umh}, suggesting that many interesting phenomena wait to be discovered. See also \cite{Gremm:1999su,Bergman:1999na,Gukov:2002er} for early, and \cite{Braun:2018joh, Eckhard:2018raj} for recent,  work on $\CN=1$  theories in the string theory context.

Starting from well known $\CN>1$ dualities known as mirror symmetries \cite{Intriligator:1996ex,Aharony:1997bx, Kapustin:1999ha}, we find dualities relating $\CN=1$ sQED with $2$ flavors to Gross-Neveu-Yukawa models. The latter theories have no gauge interactions, so they are much simpler than a QED to analyze. When they are supersymmetric they are usually called Wess-Zumino models.

We focus on theories with manifest ultraviolet (UV) symmetry $\sutf \times \uot$. Here $\sutf$ rotates the two flavors, while $\uot$ is the ``topological'' symmetry that shifts the dual photon. The monopole operators are charged under both factors, while mesonic operators (polynomial in the elementary fields) have zero topological charge. We consider four different cases, distinguished by the amount of gauge-singlet fields. These gauge-singlet fields enter the superpotential multiplying the quadratic mesons and can be in a $\sutf$-triplet or in a $\sutf$-singlet.%
\footnote{These models are in the class of $\CN=1$ gauge theories considered in \cite{Benini:2018umh}. The dualities found in  \cite{Benini:2018umh} indeed include on one side of the duality gauge-singlets in the adjoint plus singlet of the $SU(N_\text{flavor})$ symmetry. The examples of this paper however lie outside the naive range of validity of the dualities of \cite{Benini:2018umh}.}
One model has enhanced $\CN=4$ supersymmetry, and its well-known dual description in terms of free superfields is our starting point. Another model has $\CN=2$ SUSY and displays $SU(3)$ symmetry enhancement in the infrared, as recently pointed out in \cite{Gang:2017lsr, Gang:2018wek}. The other two models are genuinely $\CN=1$: one of them exhibits IR symmetry enhancement to $O(4)$, while the other (with no gauge singlets and $\CW=0$) shows no evidence of symmetry enhancement.

The enhanced symmetries are simple UV symmetries on the Wess-Zumino side, but they act non trivially on the sQED side: mesonic and monopoles operators combine into irreducible representations of the IR symmetry.

These features closely resemble the conjectured behavior of non-supersymmetric QED with 2 scalars or 2 fermions: using particle-vortex and bosonization dualities such theories can be argued to be self-dual, and it has been proposed that the $\sutf \times \uot$ UV symmetry enhances in the IR to $O(4)$ \cite{Xu:2015lxa, Hsin:2016blu, Benini:2017dus} or $SO(5)$ \cite{Wang:2017txt}. These fixed points are supposed to describe interesting quantum phase transitions. One important question is if they are stable, \ie{} if there are deformations of the theories (on top of the mass terms, which are tuned to zero) which preserve the UV symmetry and are relevant in the IR. Such a deformation would destabilize the fixed point and rule out the possibility that the CFT with enhanced symmetry describes a physical quantum phase transition.

In our cases we can take advantage of the duality with the Wess-Zumino models, which are much simpler to analyze than a gauge theory. As we did in \cite{Benini:2018umh}, we study the Wess-Zumino models in the $D=4-\varepsilon$ expansion, which is not a particularly accurate technique, but it is good enough to learn qualitative features, like whether an operator is relevant in the IR or not. We find that in all four cases the $\sutf \times \uot$ preserving deformations (basically a quartic operator which is the square of the quadratic $\sutf$ singlet meson) are irrelevant. So these $\CN=1$ fixed points are stable.

On the other hand, we find that there are relevant superpotential deformations that break the $\sutf \times \uot$ symmetry. For instance in the model with $\CW=0$ we can turn on a quartic superpotential in the isospin-2 representation of $\sutf$. Studying these $\CN=1$ deformations is certainly interesting, but goes beyond the scope of this paper \cite{toappear1}.

We also construct a model of sQED with $4$ flavors and $\CN=2$ SUSY that displays enhanced symmetry in the infrared: the topological symmetry combine with an $SO(4)$ flavor symmetry to form an $SO(6)$. From this model it is possible to construct RG flows along which some symmetry enhancement is preserved, and for instance land on the sQED with $2$ flavors and enhanced $SU(3)$ symmetry.

\subsection*{Organization of the paper}

In each Section from \ref{sec:n=4} to \ref{sec:su3} we consider a different sQED with two flavors and its dual Wess-Zumino model. We describe the superpotential, the mapping of the basic operators (which are $4$ mesonic operators and $4$ monopoles) across the duality. We check that the dual pairs have the same massive deformations, and in one case show how to deform the duality to the duality for $\CN=1$ sQED with just one flavor found in \cite{Benini:2018umh}.

In Section \ref{sec:4flav} we study the theory with $4$ flavors, argue for a self-duality, which is obtained combining Aharony duality \cite{Aharony:1997gp} with  mirror symmetry. We also check symmetry enhancement computing the superconformal index.

\paragraph{Note added.} While this work was in preparation, \cite{Gaiotto:2018yjh} appeared, which has some overlap with our paper.

\section{\matht{\CN\!=\!4} mirror symmetry for sQED with 2 flavors}
\label{sec:n=4}

We start setting out our notation and reviewing the well known mirror duality \cite{Intriligator:1996ex} for $\CN=4$ sQED with a charge $+1$ hypermultiplet, that is two complex $\CN=1$ scalar supermultiplets $Q_1$  and $Q_2$ with gauge charge $+1$. We will also consider some neutral matter fields, which we denote by $\Phi_x$. Each $\CN=1$ real scalar supermultiplet $A$ contains a real boson $a$ and a Majorana fermion $\psi_A$.

We use $\CN=1$ superspace notation with anticommuting coordinates $\theta$, and denote the matter superfields with capital letters:
\be
A(\theta) = a + \psi_A \theta + F_A \theta^2 \;.
\ee
In the theories we discuss, beyond the gauge interactions, there is also a cubic superpotential
\be
\CW = \frac{d_{ijk}}{6} \, P_i P_j P_k \;,
\ee
that in components leads to a quartic potential $V \sim (\partial W)^2 \sim p^4$ and cubic Yukawa terms of the form $\partial^2 W \psi \psi \sim p \, \psi_P \psi_P$. In other words, such theories can be considered as $\CN=1$ supersymmetric Gross-Neveu-Yukawa models, possibly gauged.

\paragraph{Monopole operators.}
A crucial role is played by local gauge-invariant operators $\M$ that are usually called monopole operators \cite{Borokhov:2002ib,Chester:2017vdh}. They carry non-zero charge under the topological symmetry associated to shifts of the dual photon. Since we consider a $U(1)$ gauge theory with zero effective Chern-Simons term and two charge 1 fermions, the bare Chern-Simon term is $1$. So the bare monopole $\M^q_\text{bare}$ has gauge charge $q$, where $q \in \bZ$ is the charge of the monopole under $\uot$. In order to get a gauge-invariant operator, we need to dress the bare monopole $\M^q_\text{bare}$ with $q$ charged elementary fields from $Q$ or $\Qd$, either the scalars $q$ or the fermions $\psi_Q$. Since the theory is supersymmetric, the monopoles form supermultiplets. We will mostly be concerned with the simplest monopoles $\M^{\pm 1}$: they have minimal topological charge and are dressed by the lowest modes of the matter superfields $Q$. Since we have 2 flavors, the monopoles will transform in a representation of $\sutf$,%
\footnote{The group $\sutf$ is useful---as a book-keeping device---also when the symmetry is broken by the interactions.}
which in this case is simply the two-dimensional representation.

Summarizing, the basic monopoles $\M^\pm$ of the various sQEDs with $2$ flavors organize into $4$ real supermultiplets, transforming as a complex doublet of $\sutf$ and with charge $\pm 1$ under $\uot$. Under the dualities we discuss, these two complex monopoles will always map to $2$ complex superfields in the dual Wess-Zumino models we propose. Generically the dual WZ models will also contain additional superfields with $\uot$ charge $0$.

\subsection*{\matht{\CN\!=\!4} SQED \matht{\;\leftrightarrow\;} free hyper}

The basic $\CN=4$ Abelian duality states that the $U(1)$ gauge theory with one hypermultiplet (\ie{} two complex $\CN=1$ multiplets) of charge 1 is dual in the IR to a free massless hypermultiplet \cite{Intriligator:1996ex, Kapustin:1999ha}. The gauge theory has R-symmetry $SU(2)_C \times SU(2)_H$ and a topological flavor symmetry $\uot$. 

We can write the $\CN=4$ sQED gauge theory in a way which is manifestly invariant under the antidiagonal subgroup $\sutf$ of the two R-symmetry factors, using the $\CN=1$ notation.%
\footnote{In the $\CN=2$ notation, the theory looks as follows: $U(1)_0$ with two chiral multiplets $Q$, $\Qt$ of gauge charge $+1$ and $-1$, respectively, and a neutral chiral multiplet $\Phi_c$, with superpotential $\CW_{\CN=2}=\Phi_c Q \Qt$. In this notation the Cartan subgroup of $SU(2)_C$ is visible as a $U(1)$ R-symmetry under which $Q, \Qt, \Phi_c$ have charges $0,0,2$, respectively, while the Cartan subgroup of $SU(2)_H$ is visible as a $U(1)$ R-symmetry under which they have charges $1, 1, 0$, respectively. Their difference is an axial flavor symmetry. In the $\cN=1$ notation there is one extra real singlet $\Phi_r$ from the $\cN=2$ vector multiplet, and the $\cN=1$ superpotential is
\be
\CW=  \Phi_r \big( Q Q^\dagger- \Qt \Qt^\dagger \big) + 2 \re \big( \Phi_c Q \Qt \big) \;.
\ee
Renaming the flavor variables $Q\rightarrow Q_1$, $\Qt \rightarrow \Qd_2$ and grouping the three real gauge-singlet superfields $\Phi_r, \Phi_c$ in a triplet $\Phi_I$, we get the presentation \eqref{N=4duality}. In this presentation $\sutf$ is manifest, with $Q_{\alpha=1,2}$ forming a doublet and $\Phi_{I=1,2,3}$ forming a triplet. Its Cartan subgroup is the axial flavor symmetry mentioned above.}
Then the duality can be stated as:
\be
\label{N=4duality}
\ba{ccc}
\ba{c}U(1)_0 \text{ with 2 flavors } Q_{\alpha=1,2} \\ \text{and 3 real fields $\Phi_{I=1,2,3}$} \\
 \CW=  \Phi_I \, Q_\a (\sigma_I)_{\a\b} \Qd_\b \ea 
    &\quad \Longleftrightarrow \quad & \quad
\ba{c} \text{Free theory of}\\
 \text{$2$ complex fields } M_{\alpha=1,2} \\
 \CW=0 \;.\ea 
   \ea
\ee
Here $\sigma_I$ are the three Pauli matrices. With a slight abuse of notation, sometimes we use $\alpha=+,-$ interchangeably with $\alpha=1,2$. The low-lying superconformal primary operators are mapped according to
\be
\label{N=4map}\ba{ccccc}
&&&&\quad \Delta\\
 \left\{ \ba{c} \M_\a \\ \Phi_I   \\ |Q_1|^2 + |Q_2|^2 \\ Q_\a (\sigma_I)_{\a\b} Q^\dag_\b  \ea  \right\} 
            & \;\; \Longleftrightarrow \;\; & 
   \left\{ \ba{c}  M_\a  \\ M_\a (\sigma_I)_{\a\b} \Md_\b   \\  |M_+|^2 + |M_-|^2 \\  D M_\a (\sigma_I)_{\a\b} D\Md_\b \ea  \right\} & \quad & \quad
   \ba{c} \frac{1}{2} \\ 1 \\1 \\ 2 \ea
\ea \ee
Here $\M_\a$ is the complex superfield whose lowest component is the scalar gauge-invariant monopole operator $\M  \, \psi_{Q_\alpha}^\dag$. The operators $\M_\a \leftrightarrow M_\a$ have charge 1 under $\uot$.
In the last column we have indicated the dimensions of the operators. The last line involves $\CN=1$ SUSY descendants of $\phi_I$. Notice that the duality gives us the relation $\sum_I \Phi_I^2 \leftrightarrow \big( |M_+|^2 + |M_-|^2\big)^2$, which implies the quantum relation
\be
\sum\nolimits_I \Phi_I^2 = \big( |Q_1|^2 + |Q_2|^2 \big)^2
\ee
in the gauge theory. Notice also that since the RHS is free, we also know that the quartic operator $\big( |Q_1|^2 + |Q_2|^2 \big)^2$ has $\Delta=2$, so it is an irrelevant $\CN=1$ deformation.  Relevant $\CN=1$ deformations include monopole superpotentials (up to order $\M^3$) and terms linear in the $\Phi_I$'s. The relevant deformations break the UV $\sutf \times \uot$ global symmetry.


\section{\matht{\CN\!=\!1} sQED \matht{\;\leftrightarrow\;} 7-field \matht{SU(2) \times U(1)} WZ model}
\label{sec:u2}

From the duality \eqref{N=4duality} and the operator map \eqref{N=4map} we perform an $\CN=1$ flip of the three gauge-singlets $\Phi_I \leftrightarrow M_\a (\sigma_I)_{\a\b} \Md_\b$, in other words we introduce three new fields $\mu_I$ and add a superpotential
\be
\delta \CW = \mu_I \, \Phi_I \qquad\longleftrightarrow\qquad  \delta \CW= \mu_I \, M_\a (\sigma_I)_{\a\b} \Md_\b \;.
\ee
This operation does not break the $\sutf \times \uot$ flavor symmetry. On the sQED side the fields $\Phi_I$ become massive and can be integrated out, while on the other side we obtain an interacting Wess-Zumino model:
\be
\label{U12flavduality}
\ba{ccc}
\ba{c} U(1) \text{ with 2 flavors } Q_1, Q_2 \\
 \CW = 0  \ea
    &\qquad\Longleftrightarrow\quad\qquad & 
\ba{c} \text{WZ model with a real triplet $\mu_I$} \\ \text{and a complex doublet $M_\a$} \\ \CW = \mu_I \, M_\a (\sigma_I)_{\a\b} M^\dagger_\b \;.
\ea 
\ea \ee
This duality appeared in \cite{Gremm:1999su}, intepreting Hanany-Witten branes setups with $\CN=1$ susy. In \cite{Gremm:1999su} it was checked that the moduli space of vacua (which is a $3$-real dimensional cone) agrees on the two sides. The superpotential of the WZ model on the RHS can be written more explicitly as
\be
\CW= \mu_3 \big( |M_+|^2-|M_-|^2 \big) + 2 \mu_1 \re\big( M_+ M_-^\dag \big) + 2\mu_2 \im\big( M_+M_-^\dag \big) \;,
\ee
or as the determinant of a $3 \times 3$ matrix
\be \CW = \det M_{U(2)} = \det \left(
\begin{array}{ccc}
 \mu_3  & \mu _1 +i \mu_2 & M_+ \\
 \mu_1 -i \mu_2 &  -\mu _3 & M_- \\
 M_+^\dagger & M_-^\dagger & 0
\end{array}
\right) \;.
\ee   
This is the most general cubic superpotential compatible with the $\sutf \times \uot$ global symmetry. There are also $\bZ_2$ discrete symmetries: $\mathbb{Z}_2^{\CT} $ times reversal and $\mathbb{Z}_2^\cC$ charge conjugation (mapping $Q_\a \mapsto Q_\a^\dagger$).

The mapping of the basic gauge-invariant operators across the duality is now:
\be
\label{U12flavmap}
\ba{ccccc}
&&&&\Delta\\
 \left\{ \ba{c} \M_\a \\ \left(\ba{cc} |Q_2|^2- |Q_1|^2 & -2 Q_1Q_2^\dagger \\ -2 Q_2Q_1^\dagger &  |Q_1|^2- |Q_2|^2 \ea\right)   \\ |Q_1|^2 + |Q_2|^2   \ea  \right\} 
            &\Longleftrightarrow& 
   \left\{ \ba{c}  M_\a \\ \left(\ba{cc} \mu_3 & \mu_1 + i \mu_2 \\ \mu_1 - i \mu_2 &  -\mu_3 \ea\right)   \\ - 2 \sum \mu_I^2 + \sum |M_\a|^2  \ea  \right\}& \quad &
   \ba{c} \sim 0.76 \\ \sim 0.66 \\ \sim 0.66 \\ \sim 1 \ea
\ea \ee
The approximate dimensions are computed at one-loop in the $\varepsilon$-expansion, as discussed below. In the last line, the particular linear combination of two flavor symmetry singlets on the RHS is dictated by the one-loop computation and it diagonalizes the dilation operator at that order.



\subsection{Relevant and irrelevant deformations}
In the WZ model, we have computed the scaling dimension of various operators in the $D=4-\varepsilon$ expansion, as in \cite{Benini:2018umh}, using the general formulas, which can be found for instance in the appendix of \cite{Fei:2016sgs}.

The value of the coupling in $\CW= \lambda \det M_{U(2)}$ at the RG fixed point at two-loops is
\be
\frac{\lambda_*}{4 \pi \sqrt{\varepsilon}} =\frac{1}{6 \sqrt{3}}+  \frac{1}{27 \sqrt{3}} \varepsilon+O(\varepsilon^2) \;.
\ee
The scaling dimension of the elementary fields are
\bea
\Delta\big[\mu_I\big] &= 1- \frac{\varepsilon}{3}- \frac{\varepsilon^2}{108} + O(\varepsilon^3) \sim 0.66 \\
\Delta \big[ M_{\pm} \big] &=1- \frac{\varepsilon}{4}+ \frac{\varepsilon^2}{144} + O(\varepsilon^3) \sim 0.76 \;.
\eea
The $28$ quadratic operators transform as
\be
({\bf3}_0 \oplus {\bf 2}_{\pm 1})^{\otimes 2}_S =  2 \cdot {\bf1}_0 \oplus {\bf3}_0 \oplus {\bf5}_0 \oplus {\bf2}_{\pm1} \oplus {\bf4}_{\pm1}  \oplus {\bf3}_{\pm2} \;,
\ee
where we denoted by ${\bf N}_q$ an operator in the $N$-dimensional irrep of $\sutf$ with charge $q$ under the $\uot$. The ${\bf3}_0$ is a SUSY descendant of $\mu_I$, while the ${\bf2}_{\pm1}$  is a SUSY descendant of $M_\a$.

The two singlets ${\bf1}_0$ have one-loop scaling dimension
\bea
\Delta \Big[ {\textstyle 2 \sum \mu_I^2 + 3 \sum |M_\a|^2} \Big] &= 2 + \frac{\varepsilon}{3}+ O(\varepsilon^2) \sim 2.33 \\
\Delta \Big[ {\textstyle - 2 \sum \mu_I^2 + \sum |M_\a|^2} \Big] &= 2 - \varepsilon+ O(\varepsilon^2) \sim 1 \;.
\eea
Since the singlet $- 2 \sum \mu_I^2 + \sum |M_\a|^2$ has $\Delta<\frac{3}{2}$, it is possible to flip it with a free $\CN=1$ real superfield. We will study this deformation in Section \ref{sec:su3}. On the other hand, the singlet $2\sum \mu_I^2 + 3\sum |M_\a|^2$ has $\Delta>2$ and so it is an irrelevant deformation. The $\sutf$ invariant quartic term $(\sum |Q_\a|^2)^2$ is not a SUSY descendant in the sQED, and it maps to the primary singlet $2 \sum \mu_I^2 + 3 \sum |M_\a|^2$. From the duality we learn that the $\sutf$ invariant quartic superpotential is an irrelevant deformation.

Let us look at the $\sutf$ isospin-$2$ operators ${\bf5}_0$. They have scaling dimension
\be
\Delta \Big[ {\textstyle \mu_I \mu_J - \frac{\delta_{IJ}}{3} \sum \mu_I^2} \Big] = 2 - \frac23 \varepsilon + O(\varepsilon^2) \sim 1.33 \;,
\ee
therefore they are relevant. They map to the quartic $\sutf$ isospin-$2$ operators in the sQED (the $\sutf$ isospin-$1$ operators are descendants of the $\sutf$ isospin-$1$ quadratic mesons). This means that it is possible to use these 5 operators to deform the sQED with $\CW=0$.


\subsection*{Massive phases}
We can give mass to both flavors of the sQED with $\CW=0$ in two different ways.

One way is $\sutf$ invariant, and uses a $\Delta \sim 1$ operator:
\be
\label{invariant massive deformation}
\delta \CW = m \big( |Q_1|^2 + |Q_2|^2 \big) \quad\longleftrightarrow\quad \delta \CW = m \Big( {\textstyle - 2 \sum \mu_I^2 + \sum |M_\a|^2} \Big) \;.
\ee
On the sQED side, depending on the sign of $m$, we are left with the pure $\cN=1$ CS gauge theory $U(1)_{\pm 1}$, which has a trivial gapped vacuum (the gaugino is massive because of the non-zero Chern-Simons term). On the WZ side, the F-terms equations are satisfied only at a single point, where all 7 fields have vanishing VEV.%
\footnote{This is true because the two terms on the RHS of (\ref{invariant massive deformation}) have opposite sign, and one obtains the equation $4m \sum \mu_I^2 = - m \sum|M_\a|^2$. Crucially, the deformation with equal signs is irrelevant and cannot be used.}
At such a point all scalars and fermions are massive, and we are left with a trivial gapped vacuum for both signs of $m$.

Another way to give mass to the two flavors breaks $\sutf$ and uses a $\Delta \sim 0.66$ operator:
\be
\label{non-invariant massive deformation}
\delta \CW = m \big( |Q_1|^2 - |Q_2|^2 \big) \quad\longleftrightarrow\quad \delta \CW = - m \mu_3 \;.
\ee
On the sQED side, for both signs of $m$ we are left with the pure $\cN=1$ gauge theory $U(1)_0$. Dualizing the photon, we get an $S^1$ worth of vacua plus a free Majorana fermion (the gaugino). On the WZ side, for $m>0$ the complex scalar $M_+$ acquires a VEV with $|M_+|^2 = m$. This spontaneously breaks $\uot$ and gives an $S^1$ of vacua, over which one real boson and one Majorana fermion are massless. For $m<0$ the story is the same, except that $M_-$ instead of $M_+$ takes a VEV.

\subsection*{Other deformations}

We can flip the operator $|Q_1|^2 - |Q_2|^2 \leftrightarrow - \mu_3$ by means of a real scalar superfield $\Psi$; this operation explicitly breaks $\sutf$ to $U(1)_\text{flav}$. On the sQED side this corresponds to the superpotential $\CW = \Psi \big( |Q_1|^2 - |Q_2|^2 \big)$. The resulting theory is, in fact, $\cN=2$ sQED with two chiral multiplets $Q = Q_1$ and $\Qt = Q_2^\dag$ of gauge charge $+1, -1$, respectively. On the WZ side the superfields $\mu_3$ and $\Psi$ become massive and can be integrated out. One is left with the superpotential $\cW = 2 \re \big[ (\mu_1 + i \mu_2) M_+ M_-^\dag\big]$. Defining $X = \mu_1 + i \mu_2$, $Y = M_+$ and $Z = M_-^\dag$, we obtain an $\cN=2$ Wess-Zumino model of three chiral multiplets with superpotential $\cW_{\cN=2} = XYZ$. Succinctly:
\be
\ba{ccc}\label{XYZDUALITY}
\ba{c} U(1)_{0} \text{ with 2 flavors } Q_1, Q_2 \\ \text{and a (real) singlet $\Psi$} \\
 \CW = \Psi \big( |Q_1|^2 - |Q_2|^2 \big)  \ea
    &\qquad\Longleftrightarrow\qquad\qquad & 
\ba{c} \text{WZ model with} \\ \text{3 complex multiplets $X,Y,Z$} \\ \CW = 2 \re \big( XYZ \big) \;.
\ea 
\ea \ee
We have obtained the well-known $\cN=2$ Abelian mirror duality \cite{Aharony:1997bx}.

Another interesting case is to turn on both massive deformations in (\ref{invariant massive deformation}) and (\ref{non-invariant massive deformation}), with a tuning that keeps one massless flavor $Q_1 = Q$ in the sQED. The resulting theory has gauge group $U(1)_{1/2}$ and is not parity invariant in the UV, therefore we expect a superpotential term $\CW = - |Q|^4$ to be generated along the RG flow. On the WZ side the massive deformation corresponds to
\be
\delta\cW = m_1 \Big( {\textstyle -2 \sum \mu_I^2 + \sum |M_\alpha|^2} \Big) - m_2 \, \mu_3 \;.
\ee
The solution to the F-term equations is $\mu_3 = -m_2/4m_1$ while all other fields have vanishing VEV. With a tuning $m_2 = 4m_1^2$ we can arrange such that, around the vacuum, only $M_-$ and $\mu_{1,2}$ are massive, while $M_+$ is not. We are left with a WZ model of a real superfield $\mu_3 = H$ and a complex superfield $M_+ = P$, with a superpotential coupling $\cW \supset H|P|^2$. It has been shown in \cite{Benini:2018umh} that in such a WZ model the term $-H^3$ is generated by the RG flow, while $H^2$ is irrelevant. We are led to the duality
\be
\ba{ccc}
\ba{c} U(1)_{1/2} \text{ with 1 flavor } Q \\
 \CW = - |Q|^4 \ea
    &\qquad\Longleftrightarrow\qquad\qquad & 
\ba{c} \text{WZ model with a real field $H$} \\ \text{and a complex field $P$} \\ \CW = H |P|^2 - H^3 \;.
\ea 
\ea \ee
This is the duality proposed in \cite{Benini:2018umh}.

%
%
%
%


\section{\matht{\CN\!=\!1} sQED with 4 singlets \matht{\;\leftrightarrow\;} 5-field \matht{O(4)} WZ model}
\label{sec:o4}

This time we flip the meson $|Q_1|^2 + |Q_2|^2 \leftrightarrow |M_+|^2 + |M_-|^2$ in the $\cN=4$ mirror duality (\ref{N=4duality}) and (\ref{N=4map}). Such operator is a singlet of $\sutf \times \uot$. We obtain a duality with the schematic form
\be
\ba{ccc}
\ba{c} U(1)_0 \text{ with 2 flavors } Q_{\a=1,2} \\ \text{and (real) singlets } H,\, \Phi_{I=1,2,3} \\
 \CW=  \Phi_I Q_\a (\sigma_I)_{\a\b} \Qd_\b + H Q_\a Q_\a^\dag + \ldots \ea 
    & \Longleftrightarrow\quad\, &
\ba{c} O(4) \text{ WZ model with} \\ \text{1 real and 2 complex fields } H,\, M_\a \\ 
 \CW= H M_\a M_\a^\dag + \ldots \ea 
\ea \ee
where the dots stand for quantum corrections that we describe below. The operator map is
\be
\label{O4map}\ba{ccccc}
&&&&\quad\Delta\\
 \left\{ \ba{c} \M_\a \\ \Phi_I   \\ H  \ea  \right\} 
            & \qquad\Longleftrightarrow \;\qquad\qquad & 
   \left\{ \ba{c}  M_\a  \\ M_\a (\sigma_I)_{\a\b} \Md_\b   \\  H  \ea  \right\} & \quad & \quad
   \ba{c} \sim 0.62 \\ \sim 1.46 \\ \sim 0.81  \ea
\ea \ee
As we are going to describe, this duality implies an IR symmetry enhancement in the sQED:
\be
\sutf \times \uot \quad\rightarrow\quad O(4) \;.
\ee
On the WZ side, the four real superfields contained in $M_+, M_-$ transform in the vector representation of $O(4)$, while $H$ is a singlet. From the duality we infer how the operators organize on the sQED side.
   
For instance, on the WZ side there are nine quadratic operators in the symmetric traceless representation $(\rep3, \rep3)$ of $O(4)$. The three of them with charge 2 under $\uot$, namely $\big( M_+^2, M_+M_-, M_-^2 \big)$, are mapped to monopole operators $\M^{+2}$ on the sQED side, while the three of them with charge 0, namely $M_\a (\sigma_I)_{\a\b} M_\b^\dag$, are mapped to the singlets $\Phi_I$. We conclude that the 9 operators $\M^{+2}, \Phi_I, \M^{-2}$ must be degenerate in the sQED.

The full quantum superpotential for the sQED has the general form
\bea
\label{superpotential sQED O(4)}
\CW &= \a  \left[ \Phi_3 \big( |Q_1|^2 - |Q_2|^2 \big) + 2\re\Big( (\Phi_1 - i \Phi_2) Q_1 Q_2^\dag \Big) \right] + {} \\
&\quad {} + \b \, H \big( |Q_1|^2 + |Q_2|^2 \big) -\gamma \, H \sum \Phi_I^2 + \delta \, H^3 \;.
\eea
On the other hand, the full quantum superpotential for the Wess-Zumino model has the general form
\be
\CW = H \big( |M_+|^2 + |M_-|^2 \big) - \lambda \, H^3 \;.
\ee
This WZ model was studied in \cite{Benini:2018umh}: it was found that the sign of the quantum-generated cubic superpotential is negative (\ie{} $\lambda>0$). Also, the one-loop scaling dimensions reported in eqn. \eqref{O4map} were computed.

\paragraph{The flavor-singlet operators.}
On the sQED side, the first parity-even flavor-singlet scalar operators are
\be
H^2\,,\quad  \sum \Phi_I^2\,,\quad \sum |Q_\a|^2\,,\quad \ldots
\ee
A linear combination of them is a descendant of $H$. Similarly, on the WZ side the first parity-even flavor-singlet operators are
\be
H^2\,,\quad \sum |M_\a|^2\,,\quad \ldots
\ee
A linear combination of them is a descendant of $H$. The other linear combination was found in \cite{Benini:2018umh} to be irrelevant:
\be
\Delta \Big[ {\textstyle \sum |M_\a|^2 + H^2} \Big] \sim 2.23 \;.
\ee
It is also likely that all higher operators, such as $\big( \sum |M_\a|^2 \big)^2$, $H^2 \sum |M_\a|^2$ and so on, have $\Delta>2$. If that is true, all parity-even $O(4)$ invariant operators have $\Delta>2$ in the WZ model. Using the duality, it follows that all parity-even $\sutf \times \uot$ invariant deformations of the sQED
are irrelevant.

\subsection*{Massive deformations}
The most natural deformation is $O(4)$ invariant and parity-odd:
\be
\delta \CW = m \, H \;.
\ee
Recall that parity-odd deformations of the superpotential preserve parity.

Consider first the sQED side, for $m>0$. At least assuming that the signs in the superpotential at the fixed point are as in (\ref{superpotential sQED O(4)}), the triplet $\Phi_I$ acquires a VEV and breaks the $\sutf$ symmetry to $U(1)$. The vacua sit on a $S^2$:
\be
\sum \Phi_I^2 = m \;,\qquad H = Q_\alpha = 0 \;.
\ee
At each point of $S^2$, the two flavors $Q_\a$ get opposite mass and leave a free $\cN=1$ gauge theory $U(1)_0$ in the IR. The bosonic part of the theory describes a $S^1$ fibered over $S^2$, namely a NLSM with target $S^3$.

For $m<0$ we find two vacua related by spontaneously-broken parity symmetry: $\Phi_I = Q_\a= 0$ and $H = \pm\sqrt{-m}$. In each vacuum the two flavors get a mass of the same sign, leaving a pure $\CN=1$ CS gauge theory $U(1)_{\pm 1}$ with a trivial gapped vacuum.

Let us now look at the WZ side. For $m<0$ there is an $S^3$ worth of vacua, $\sum |M_\alpha|^2 = -m$ and $H=0$, from the spontaneous breaking $O(4) \to O(3)$. For $m>0$ there are two vacua sitting at $M_\a = 0$ and $H = \pm \sqrt m$, where parity is spontaneously broken. The vacua match upon mass deformations (up to an uninfluential sign redefinition of $H$).



\section{\matht{\CN\!=\!2} sQED \matht{\;\leftrightarrow\;} \matht{\CN\!=\!1} 8-field \matht{SU(3)} WZ model}
\label{sec:su3}

Starting from the dualities in the previous sections, we can obtain a duality for the $\CN=2$ theory $U(1)$ with 2 chiral multiplets of charge $+1$.%
\footnote{\label{chargesign}In 3d QED's, the notion of sign of gauge charges is meaningful only for $\CN=2$ SUSY. For $\CN>2$ the flavors always come in pairs, that in $\CN=2$ language have opposite charge. For $\CN=0,1$ it is always possible to make a field redefinition (exchange a field with its complex conjugate) that changes the sign of the charge. In the $\cN=2$ case, instead, if we write a theory with $N_+$ flavors of charge $+1$ and  $N_-$  flavors of charge $-1$ in $\CN=1$ language, there are $\CN=1$ superpotential interactions that break $SU(N_++N_-)$ to $SU(N_+) \times SU(N_-)$.}
In $\CN=1$ language, this theory reads
\be
\ba{c}
U(1) \text{ with 2 flavors } Q_1,\, Q_2 \text{ and a (real) singlet $\Phi$} \\[.5em]
\CW= \Phi \big( |Q_1|^2 + |Q_2|^2 \big) \;.
\ea
\ee
The continuos UV global symmetry is $\sutf \times \uot \times U(1)_R$.
 
\paragraph{Enhanced global symmetry.} 
This $\CN=2$ sQED is expected to have enhanced $SU(3)$ global symmetry in the IR, as pointed out in \cite{Gang:2017lsr, Gang:2018wek}, using geometrical features of the $3d/3d$ correspondence.

It is possible to argue for the enhanced global symmetry in the following way. Using the duality between a free chiral multiplet and $\CN=2$ $U(1)_{1/2}$ sQED with one flavor:
\be
\ba{ccc}
\ba{c} U(1)_{1/2} \text{ with 1 flavor } Q \\
 \CW = \Psi Q\Qd - \frac{1}{2}\Psi^2 \ea
    &\qquad\Longleftrightarrow\qquad\qquad & 
\ba{c} \text{Free complex superfield $P$} \\ \CW = 0 \;,
\ea 
\ea \ee
 it is possible to show a self-duality of the $\CN=2$ gauge theory $U(1)$ with $2$ flavors of charge $+1$. This self-duality exchanges $2$ mesons with $2$ monopoles:
\be
\{Q_1\Qd_2 \,,\, Q_2\Qd_1\} \quad\longleftrightarrow\quad \{ \M^+Q_1 \,,\, \M^-\Qd_1 \} \;.
\ee
Since the $2$ mesons transform in the $\rep{3}_0$ of $\sutf \times \uot$ while the $2$ monopoles transform in $\rep{2}_{\pm 1}$, it must be that the UV flavor symmetry $\sutf \times \uot$ enhances in the IR, in this case to $SU(3)$.%
\footnote{The same self-duality can be argued for the $\CN=2$ sQED with $1$ flavor of charge $+1$ and $1$ flavor of charge $-1$ of \eqref{XYZDUALITY}. In this case the self-duality exchanges the complex meson $Q\Qt$ (which is in the chiral ring) with a BPS chiral monopole. Together with parity, which exchanges monopoles with anti-monopoles, it implies that there is a quantum IR $S_3$ symmetry. This $S_3$ is of course manifest in the dual $\CN=2$ $XYZ$ Wess-Zumino model.} 

In the UV there are four classical flavor currents, corresponding to $\sutf \times \uot$. In the IR there must be four extra accidental currents from monopole operators. Since with $\CN=2$ supersymmetry each current supermultiplet starts with a scalar operator of dimension $\Delta=1$, we expect eight such scalar operators in total. Four of them are $\big\{Q_\a (\sigma_I)_{\a\b} \Qd_\b \,,\, \Phi \big\}$ (a triplet and a singlet of $\sutf$, respectively), while the other four are the scalar monopole operators $\{\M^\pm \psi_{Q_\a}\}$ (two with topological charge $+1$ and two with $-1$). All the $8$ basic operators must form a ${\bf 8}$ of $SU(3)$ and have $\Delta=1$.
 
\subsection*{A dual \matht{\CN\!=\!1} WZ model with manifest \matht{SU(3)} symmetry}
In this paper we are interested in studying the symmetry enhancement using the duality with the Wess-Zumino model, which will make the $SU(3)$ manifest (even though it hides the extended SUSY).

Starting from the duality  \eqref{U12flavduality} and its operator map \eqref{U12flavmap}, we can find the dual $\CN=1$ WZ model.\footnote{We could as well start from the $\CN=4$ duality (flipping the 3 singlets $\Phi_I$ and the meson), or from the duality of Section \ref{sec:o4} (flipping all 4 singlet fields).} We flip the mesonic singlet 
\be\label{flippinflow}
|Q_1|^2 + |Q_2|^2 \qquad\longleftrightarrow\qquad  {\textstyle - 2 \sum \mu_I^2 + \sum |M_\a|^2} \;.
\ee
On the LHS we obtain the $\CN=2$ gauge theory $U(1)$ with two flavors of charge $+1$ ($Q_1=Q$, $Q_2=\Qt^\dag$), while on the RHS we obtain a cubic WZ model with a total of eight real superfields:
\be \ba{ccc}
\ba{c} U(1) \text{ with 2 flavors } Q_1,\, Q_2 \\ \text{and a (real) singlet $\Phi$} \\
 \CW= \Phi \big( |Q_1|^2 + |Q_2|^2 \big)  \ea 
    &\qquad\Longleftrightarrow\qquad\quad &
\ba{c} \text{WZ model }\\    
    \CW = \Phi \big( {-}2 \sum \mu_I^2 + \sum |M_\a|^2 \big) \\
    {} + \mu_I M_\a (\sigma_I)_{\a\b} M^\dag_\b + \ldots \ea 
   \ea \ee
On the sQED side no additional superpotential terms can be generated, due to the enhanced $\CN=2$ SUSY. On the WZ side, instead, we expect a cubic superpotential term $\Phi^3$ to be generated quantum mechanically: this is the only other cubic $\sutf \times \uot$ singlet which is also parity-odd. The most general cubic superpotential that respects the $\sutf \times \uot$ symmetry is then
\bea
\label{u2W}
\CW &= \a  \left[ \mu_3 \big( |M_+|^2 - |M_-|^2 \big) + 2\re\Big( (\mu_1 - i \mu_2) M_+ M_-^\dag \Big) \right] + {} \\
&\quad {} + \frac{\b}{\sqrt3} \, \Phi \big( |M_+|^2 + |M_-|^2 \big) - \frac{2\gamma}{\sqrt3} \, \Phi \sum \mu_I^2 + \frac{2\delta}{3\sqrt3} \, \Phi^3 \;.
\eea
The reason we chose such normalization of the four couplings is that if we organize the 8 real superfields in a $3 \times 3$ matrix as
\be M_{SU(3)} = \left(
\begin{array}{ccc}
 \mu _3\!-\!\frac{\Phi}{\sqrt{3}}  & \mu _1+i \mu _2 & M_+ \\
 \mu _1-i \mu _2 &  -\mu _3\!-\frac{\Phi}{\sqrt{3}} & M_- \\
 M_+^\dagger & M_-^\dagger & 2 \frac{\Phi}{\sqrt{3}}  \\
\end{array}
\right) \;,
\ee
then with $\a=\b=\gamma=\delta$ the superpotential becomes simply
\be
\CW = \alpha \det M_{SU(3)}
\ee
and the Wess-Zumino model enjoys $SU(3)$ global symmetry, since $M_{SU(3)}$ transforms in the adjoint representation of $SU(3)$.

The operator map is very similar to the ones for the other dualities:
\be
\label{su3map}
\ba{ccc}
 \left\{ \ba{c} \M_\alpha \\ \left(\ba{cc} |Q_2|^2- |Q_1|^2 & -2 Q_1Q_2^\dagger \\ -2 Q_2Q_1^\dagger &  |Q_1|^2- |Q_2|^2 \ea\right)   \\ \Phi   \ea  \right\} 
            &\Longleftrightarrow& 
   \left\{ \ba{c}  M_\alpha  \\ \left(\ba{cc} \mu_3 & \mu_1 + i \mu_2 \\ \mu_1 - i \mu_2 &  -\mu_3 \ea\right)   \\ \Phi  \ea  \right\}
   \ea \ee
At the $SU(3)$ invariant point, the scaling dimension of all 8 fields in $M_{SU(3)} $ is expected to be $\Delta=1$ (as imposed by $\cN=2$ supersymmetry).

Parity can be taken to act as $M_{SU(3)} \rightarrow - M_{SU(3)} $, while charge conjugation acts as $M_{SU(3)} \rightarrow M_{SU(3)}^* = M_{SU(3)}^T$.
Notice that the $U(1)$ R-symmetry and the $\CN=2$ supersymmetry are not visible in the UV in the WZ model.

\subsection*{Analisys of the Wess-Zumino model in the \matht{D\!=\!4-\varepsilon} expansion}

We analyze the $\CN=1$ Wess-Zumino model \eqref{u2W} perturbatively using the $D=4-\varepsilon$ expansion, as in \cite{Benini:2018umh}.

The one-loop beta-functions for the four couplings in \eqref{u2W} are:
\bea
\b_{\a} &= -\a \varepsilon + \frac{\a}{3 \pi ^2} \left(9 \a^2+2 \b^2 - 4 \beta \gamma+2 \gamma^2 \right)\\
\b_{\b} &= -\frac{\b}{\sqrt{3}} \varepsilon + \frac{1}{3 \sqrt{3} \pi ^2} \big(18 \a^2 (\b-\gamma)+\b (3 \b^2+2 \b \delta+3 \gamma^2+\delta^2 ) \big)\\
\b_{\gamma} &= -\frac{\gamma}{\sqrt{3}} \varepsilon + \frac{1}{3 \sqrt{3} \pi ^2} \big(6 \a^2 (\gamma - \b)+\gamma (\b^2+11 \gamma^2 - 4 \gamma \delta+\delta^2 ) \big)\\
\b_{\delta} &=  -2 \frac{\delta}{\sqrt{3}} \varepsilon +  \frac{2}{3 \sqrt{3} \pi ^2} \left(2 \b^3+3 \b^2 \delta - 12 \gamma^3+9 \gamma^2 \delta+7 \delta^3 \right) \;.
\eea
We look for critical points where the beta-functions vanish.
Notice that there is no solution with $\delta=0$ and $\b \neq 0$ or  $\gamma \neq 0$: as expected the term $\Phi^3$ is generated. Modulo $\a \rightarrow -\a$ and $\delta \rightarrow - \delta$ (which can be obtained by field redefinitions), the solutions of the beta-function equations (with all couplings turned on) for the four rescaled couplings
\be
\frac{1}{4 \pi \sqrt{\varepsilon}}\left\{ \a, \b , \gamma, \delta \right\}
\ee
are:
\bea
\frac{1}{4 \sqrt{3}}\{1,1,1,1\} \,\quad & \{0.032, -0.279, 0.020, 0.171 \} \\
\{0.051, -0.245, 0.041,0.161 \} \,\quad & \{0.052, -0.156 , 0.130, 0.159\} \;.
\eea
It can be checked that the first fixed point---with enhanced $SU(3)$ symmetry---and the last one are stable. At the other two fixed points the direction parametrized by $\gamma$ is unstable. This implies that the $\sutf \times \uot$ global symmetry enhances in the IR
to $SU(3)$.

We have computed the scaling dimension of the 8 elementary fields at the $SU(3)$ invariant point at two-loops,%
\footnote{At the $SU(3)$ invariant point, where $\CW = \alpha \det M_{SU(3)}$, the two-loop beta-function and scaling dimensions are 
\be
\b_{\a}= - \a \varepsilon  +\frac{48}{(4 \pi )^2}\a^3 - \frac{4864 }{3(4 \pi )^4} \a^5 \;,\qquad\qquad
\Delta[{\bf 8}] = 1- \frac{\varepsilon}{2}  +\frac{5 }{6 \pi ^2}\a^2 - \frac{25 }{18 \pi ^4}\a^4 \;.
\ee
Notice that the beta-function equation should be solved perturbatively in $\varepsilon$.}
finding
\be
\Delta[{\bf 8}] =1- \frac{\varepsilon}{2}+\frac{5 \varepsilon}{18}+ \frac{10 \varepsilon^2}{243} +  O(\varepsilon^3) \sim 0.82 \;.
\ee
Notice that this result is quite far from the exact scaling dimension $\Delta=1$ that is implied by the duality. This means that the $\varepsilon$ expansion has poor accuracy and can only be used to infer qualitative features of the IR CFT.

The 36 quadratic operators transform under $SU(3)$ as
\be
({\bf 8} \oplus {\bf 8})^{\otimes 2}_S = {\bf1} \oplus {\bf 8} \oplus {\bf 27} \;.
\ee
The ${\bf 8}$ is a SUSY descendant of the elementary fields.

The other two representations have one-loop scaling dimension
\be
\Delta[{\bf 1}] = 2 - \varepsilon + \frac{5\varepsilon}{3}+ O(\varepsilon^2) \sim 2.33
\ee
and
\be
\Delta[{\bf 27}] = 2 - \varepsilon + \frac{7\varepsilon}{9} + O(\varepsilon^2) \sim 1.77 \;.
\ee
Also in this theory the flavor singlet is irrelevant, so the quartic $\sutf$ invariant superpotential on sQED side cannot be turned on.

\subsection*{Massive phases}

We consider the $SU(3)\rightarrow SU(2) \times U(1)$ breaking deformation 
\be
\delta \CW = m \, \Phi \;.
\ee
The discussion is parallel to the case of the previous section, since the superpotentials are the same: what changes is on which side the gauge field sits.

On the sQED side, for $m>0$ the quarks $Q_\a$ acquire a VEV which sits on an $S^3$: $\sum |Q_\a|^2=m$ and $\Phi=0$. Quotienting by the gauge action, we obtain an $S^2$ worth of vacua. For $m<0$ we need to include one-loop quantum effects, since we are going to move along the Coulomb branch of an $\CN=2$ gauge theory. When $\Phi$ acquires a VEV, the quarks get masses of the same sign and an effective Chern-Simons term with level $\mathrm{sign}(\Phi)$ as well as its supersymmetric partner $\delta \CW = \mathrm{sign}(\Phi) \Phi^2$ are generated. This happens both for $\Phi=m$ and $\Phi=-m$. Concluding, for $m<0$ there are two gapped vacua.

On the WZ side, for $m>0$ the triplet $\mu_I$ acquires a VEV: the vacua sit on a $S^2$ parametrized by $\sum \mu_I^2=m$ and $M_\a = \Phi=0$.  For $m<0$ there are two vacua related by spontaneously-broken parity symmetry: $\Phi = \pm m$ and $\mu_I = M_\a= 0$.
The vacua match upon mass deformations.


\section{\matht{SO(6)} enhancement in \matht{\CN\!=\!2} sQED with \matht{2\!+\!2} flavors}
\label{sec:4flav}

In this section we consider the gauge theory $U(1)$ with $4$ flavors $Q_\a, \Qt_\b$. In the case of $\CN=4$ SUSY, that is $\CN=4$ $U(1)$ gauge theory with 2 hypermultiplets whose UV global symmetry is $SU(2) \times \uot \times SO(4)_R$, it is well known that the theory is self-dual under mirror symmetry \cite{Intriligator:1996ex}. The self-duality exchanges monopoles with mesons, and implies an IR symmetry enhancement to $SU(2) \times SU(2) \times SO(4)_R$.

It is conceivable that similar self-dualities exist for QED with $4$ flavors and less supersymmetry. Such phenomena might also have interesting implication for quantum phases transitions. Here we present an example with $\CN=2$ SUSY, that to the best of our knowledge has not been discussed in the literature.%
\footnote{We are indebted with Sara Pasquetti for stimulating discussions about this and related topics.}
In this section we use $\CN=2$ notation and all superpotentials are complex $\CN=2$ superpotentials.

We consider the $\CN=2$ sQED theory with 2 flavors $Q_\alpha$ of charge $+1$, 2 flavors $\Qt_\beta$ of charge $-1$ and 4 gauge-singlet chiral fields $\eta_{\a\b}$.%
\footnote{See also Footnote \ref{chargesign}.}
The complex $\CN=2$ superpotential
\be
\label{so6theory}
\CW_{\CN=2}=  \sum_{\a,\b=1}^2 \eta_{\a\b} Q_\a \Qt_\b
\ee
is manifestly $SU(2)_L \times SU(2)_R$ invariant. The continuos UV global symmetry of the theory is 
\be
SU(2)_L \times SU(2)_R \times U(1)_{a} \times \uot \times U(1)_R \;.
\ee 
Here $SU(2)_L$ and  $SU(2)_R$ rotate the quarks $Q_\a$ and $\Qt_\b$, respectively, while $\eta_{\a\b}$ transform as a bifundamental of $SU(2)_L \times SU(2)_R$. 

The chiral ring is generated by the four gauge singlets $\eta_{\a\b}$ and the two SUSY chiral monopoles $\M^\pm$ (which are singlets under the $SU(2)^2$). The holomorphic mesons $Q_\a \Qt_\b$ are set to zero by the F-terms of $\eta_{\a\b}$. We will see that these operators satisfy a single quadratic quantum relation. 

Normalizing the $U(1)_a \times U(1)_R$ charges of the fundamental flavors as $(1, r)$, all six chiral ring generators have $U(1)_a \times U(1)_R$ charges $(-2,2-2r)$. The superconformal R-charge $\mathbf{r}$ of the fundamental flavors can be determined to great accuracy numerically using $\CZ_{S^3}$-extremization:
\be
\label{scR}
\mathbf{r} = 0.6696\ldots  \;.
\ee

\subsection{Self-duality and symmetry enhancement}

We want to argue for a self-duality of the theory in \eqref{so6theory}. We focus on the chiral ring generators. Let us recall that the 3d $\CN=2$ theory $U(1)$ with $2+2$ flavors and $\CW=0$ satisfies two different IR dualities. 

The first one is Aharony duality \cite{Aharony:1997gp}:
  \be \label{x} \ba{ccc}
\ba{c} U(1) \text{ with 2 flavors }  Q_i,\, \Qt_i \\
   \CW=  0 \ea 
    &\qquad\Longleftrightarrow\qquad\qquad & 
\ba{c} U(1) \text{ with 2 flavors } P_i,\, \Pt_i \\
   \CW = \sum_{i,j} \mu_{ij} \Pt_i P_j + \sum_\pm \mu_{\pm}\M^{\pm} \ea 
      \ea \ee
The six chiral ring generators map as
\be \ba{ccc}
 \left\{\ba{c}  \Qt_1Q_1 \,,\,  \Qt_2 Q_2 \\  \Qt_1Q_2 \,,\, \Qt_2 Q_1 \\ \M^\pm \ea  \right\}&\Longleftrightarrow&  
  \left\{\ba{c}  \mu_{11} \,,\, \mu_{22} \\ \mu_{12} \,,\, \mu_{21} \\ \mu_\pm \ea  \right\}
     \ea \;.
\ee
The second duality is an $\CN=2$ version \cite{Aharony:1997bx} of mirror symmetry \cite{Intriligator:1996ex}:
  \be \label{x} \ba{ccc}
\ba{c} U(1) \text{ with 2 flavors } Q_i,\, \Qt_i \\
   \CW=  0 \ea 
    &\qquad\Longleftrightarrow\qquad\qquad & 
\ba{c} U(1) \text{ with 2 flavors } R_i,\, \Rt_i \\
   \CW =  \sum_{i=1}^2 \phi_{i} \Rt_i R_i \ea 
      \ea \ee
The six chiral ring generators map as
\be \ba{ccc}
 \left\{\ba{c}  \Qt_1Q_1 \,,\, \Qt_2 Q_2 \\  \Qt_1Q_2 \,,\, \Qt_2 Q_1 \\ \M^\pm \ea  \right\}&\Longleftrightarrow&  
  \left\{\ba{c}  \phi_1\,,\, \phi_2 \\ \M^\pm \\ \tilde{R}_1R_2 \,,\, \tilde{R}_2 R_1 \ea  \right\}
     \ea \;.
\ee

We can apply the dualities above to the theory in \eqref{so6theory}. Basically, we start from either one of the above dualities and flip on both sides the right operators, using the mapping of chiral ring generators. Starting from \eqref{so6theory}, applying Aharony duality horizontally and mirror symmetry vertically, we obtain the following duality web:
\be\ba{ccc}
\ba{c} U(1)\, \text{w/} \, 2 \, \textrm{flavors} \,  Q_i,\Qt_i \\
   \CW=  \sum_{\a,\b=1}^2 \eta_{\a\b} Q_\a \Qt_\b \ea 
    &\Longleftrightarrow& 
\ba{c} U(1)\, \text{w/} \, 2 \, \textrm{flavors} \,  P_i,\Pt_i \\
   \CW =  \sum_\pm \mu_{\pm}\M^{\pm} \ea           \\
\Big\Updownarrow && \Big\Updownarrow \\
\ba{c} U(1)\, \text{w/} \, 2 \, \textrm{flavors} \,  R_i,\Rt_i \\
   \CW=  \sum_{\pm} \phi_\pm \M^\pm \ea 
    &\Longleftrightarrow& 
\ba{c} U(1)\, \text{w/} \, 2 \, \textrm{flavors} \,  S_i,\tilde{S}_i \\
   \CW = \sum_i \rho_{i} S_i\tilde{S}_i + \rho_{+} S_1\tilde{S}_2 + \rho_{-}S_2\tilde{S}_1 \ea 
\ea
\ee
Looking at the top-left and bottom-right theories, we recognize that---composing Aharony duality and mirror symmetry---we get a self-duality of the theory in \eqref{so6theory}. We can think of this as a quantum $\mathbb{Z}_2^\text{dual}$ symmetry emerging in the infrared.%
\footnote{Of course also the theories sitting at the top-right and bottom-left corners, with only two chiral gauge-singlets fields, are self-dual and enjoy the enhanced symmetry. They are the same in the IR.}
Under self-duality, the six chiral ring generators map as 
\be\ba{ccc}
\left\{\ba{c}  \eta_{11} \,,\, \eta_{22} \\  \eta_{12} \,,\, \eta_{21} \\ \M^\pm \ea  \right\}
    &\Longleftrightarrow& 
 \left\{\ba{c}  Q_1\Qt_1 \,,\, Q_2\Qt_2 \\ Q_1\Qt_2 \,,\, Q_2\Qt_1 \\ \mu_\pm \ea  \right\}  \\
\Big\Updownarrow && \Big\Updownarrow \\
 \left\{\ba{c}  R_1\tilde{R}_{1} \,,\, R_2\tilde{R}_{2} \\ \phi_\pm  \\ R_1\tilde{R}_{2} \,,\, R_2\tilde{R}_{1}  \ea  \right\}
    &\Longleftrightarrow& 
 \left\{\ba{c}  \rho_{1} \,,\, \rho_{2} \\ \M^\pm  \\ \rho\pm \ea  \right\}
 \ea \ee
From the final diagram we learn that $\mathbb{Z}_2^\text{dual}$ acts on the six chiral ring generators as
\be \ba{ccc}
 \left\{\ba{c}  \eta_{11} \,,\, \eta_{22} \\  \eta_{12} \,,\, \eta_{21} \\ \M^\pm \ea  \right\}&\Longleftrightarrow&  
  \left\{\ba{c}  \eta_{11} \,,\, \eta_{22} \\ \M^\pm  \\ \eta_{12} \,,\, \eta_{21} \ea  \right\}
     \ea \;.
\ee
Since the self-duality exchanges the gauge singlets $\eta_{12}$ and $\eta_{21}$ (which are part of the representation $({\bf 2}, {\bf 2})_0$ of $SU(2)_L \times SU(2)_R \times \uot$) with the monopoles $\M^\pm$ (which are in the $({\bf 1}, {\bf 1})_{\pm 1}$), it must be that the UV global symmetry $SU(2)_L \times SU(2)_R \times \uot$ enhances in the IR to a bigger rank 3 group. Besides, $\bZ_2^\text{dual}$ is part of the Weyl group of the enhanced symmetry, which in this case must be $SO(6)_\text{en}$. Thus, the combination of Aharony duality and mirror symmetry implies that the IR symmetry of the theory is enhanced to
$$
SO(6)_\text{en} \times U(1)_a \times U(1)_R \;.
$$

The adjoint of $SO(6)_\text{en}$ decomposes into irreps of $SU(2)_L \times SU(2)_R \times \uot$ as follows:
\be
{\bf 15} \rightarrow ({\bf 1},{\bf 3})_{0} \oplus ({\bf 3},{\bf 1})_{0}  \oplus ({\bf 1},{\bf 1})_{0} \oplus ({\bf 2},{\bf 2})_{\pm 1} \;.
\ee
The last term represents eight emergent current multiplets, which are monopoles with topological charge $\pm 1$. They are $\CN=2$ real multiplets, whose bottom component is a scalar with $\Delta=1$ and zero $R$-charge.%
\footnote{These scalar monopoles arise from dressing the bare charge $\pm1$ monopoles with $2$ fermionic zero-modes, as required by gauge invariance (recall that the bare Chern-Simon level is $-N_f/2$). Since the two fermions must be antisymmetrized, we have in total $6+6$ states. $4$ of them are chiral/antichiral BPS monopoles, the other $8$ are in the $8$ conserved supermultiplets which emerge in the IR.}

\paragraph{Massive deformation to the \matht{SU(3)} sQED with 2 flavors.}
The $\CN=2$ ``real mas'' operator in the Cartan of $SU(2)_R$,
\be
|Q_1|^2 - |Q_2|^2 - \sum_{\b=1}^{2}( |\eta_{1\b}|^2 - |\eta_{2\b}|^2) \;,
\ee
sits in the adjoint representation of $SO(6)_\text{en}$, is neutral under $SU(2)_L \times \uot$, and breaks $SU(2)_R$ to $U(1)_b$. 
Therefore, turning this deformation on breaks $SO(6)_\text{en}$ to a subgroup $SO(4)_\text{en} \times U(1)_b$ and triggers an RG flow along which the topological symmetry is always enhanced to a non-Abelian group. In the IR the flavors $\Qt$ and the $4$ complex singlets $\eta$ are massive: we are left with sQED with $2$ flavors $Q_1, Q_2$ and zero Chern-Simons level (because the two massive flavors have opposite mass), discussed in Section~\ref{sec:su3}. In the IR only an $SU(2)_\text{en}$ subgroup of $SO(4)_\text{en} \times U(1)_b$ acts, while the rest of the group acts trivially. The factor $U(1)_a$ enhances to $\sutf$. The two IR symmetries $SU(2)_\text{en}$ and $\sutf$ do not commute and combine into an $SU(3)$ symmetry. This RG flow presents a different perspective on the symmetry enhancement of the theory considered in Section~\ref{sec:su3}.

\subsection{Superconformal index, chiral ring and moduli space of vacua}

As a further check of the claimed symmetry enhancement, let us compute the superconformal index of the theory. Defining
\be
f_\Delta[s,x,t] = \frac{t x^{|m|+\Delta}-t^{-1} x^{|m|+2-\Delta}}{1-x^2} \;,
\ee
the single-letter partition function for an Abelian gauge theory with $2+2$ flavors and $4$ singlets is
\be
f_\text{s.l.}(z,a,b,t_a) = f_{\bf r}[s/2, x, a^{\pm 1} t_a z] +f_{\bf r}[s/2, x, b^{\pm 1} t_a / z] + f_{2-2{\bf r}}[ 0 , x, a^{\pm 1} b^{\pm 1} t_a^{-2} ] \;.
\ee
Here $z, a, b, t_a$ are the fugacities for $U(1)_\text{gauge} \times SU(2)_L \times SU(2)_R \times U(1)_{a}$, while $\mathbf{r}$ is the R-charge of the flavors, whose superconformal value is given in \eqref{scR}.

The superconformal index that includes the fugacity $t$ for $\uot$ \cite{Imamura:2011su} is a sum over all monopole sectors:%
\footnote{The Plethystic Exponential $PE$ of a function $f(t)$ such that $f(0)=0$ is defined as
\be
PE\big[ f(t) \big] = \text{Exp}\left(\sum_{n=1}^{\infty} \frac{1}{n} f(t^n) \right) \;.\ee
}
\be
\text{SC-I}(t,a,b,t_a) = \sum_{m=-\infty}^{+\infty} t^{m}\int \frac{dz}{z} \, x^{2-2{\bf r}} \, t_a^{-|m|} \; PE\big[ f_\text{s.l.}(z,a,b,t_a) \big] \;.
\ee
It can be checked that it indeed admits an expansion at small $x$ in terms of characters $\chi_{\so(6)}$ of $SO(6)_\text{en}$ :
\begin{multline}
\label{SCIexpan}
\text{SC-I} = 1+ x^{2-2{\bf r}} \, t_a^{-2} \, \chi_{\so(6)}[{\bf 6}] +  x^{4-4{\bf r}} \, t_a^{-4} \, \chi_{\so(6)}[{\bf 20'}] \\
{} +  x^{3-3{\bf r}} \, t_a^{-6} \, \chi_{\so(6)}[{\bf 50}] - x^2 \big( \chi_{\so(6)}[{\bf 15}] + 1 \big) + O(x^2) \;,
\end{multline}
where 
\be
\chi_{\so(6)}[ {\bf 6} ] = a b+\frac{a}{b}+\frac{1}{a b}+\frac{b}{a}+t+\frac{1}{t} \;.
\ee
The first three contributions in \eqref{SCIexpan} are chiral ring operators (linear, quadratic, cubic in the generators), while the term proportional to $-x^2$ is the contribution of fermionic operators sitting in the IR conserved current multiplets. We indeed see that the adjoint of $SO(6)_\text{en}$ appears there.

Let us close with a description of the chiral ring and the moduli space of vacua. From the expansion \eqref{SCIexpan} we can see that the six chiral ring generators satisfy a quadratic equation, which is  an $SO(6)$ invariant quantum relation for the monopoles:
\be
\M^+ \M^- = \eta_{\a\b} \, \eta_{\gamma \delta} \, \epsilon^{\a\gamma} \, \epsilon^{\b \delta}=2(\eta_{11}\eta_{22}-\eta_{12}\eta_{21}) \;.
\ee
The moduli space of vacua of the gauge theory is the $5$-complex dimensional cone defined by one $SO(6)$ invariant quadratic equation in $6$ variables. So in particular the moduli space of vacua is a complete intersection. This implies that the chiral ring Hilbert Series \cite{Benvenuti:2006qr} is a simple Plethystic Exponential:
\be
\mathcal{HS} = PE \big[\chi_{\so(6)}[ {\bf 6} ]t - t^2 \big] = 1+ \chi_{\so(6)}[{\bf 6}] t + \chi_{\so(6)}[{\bf 20'}] t^2 + \chi_{\so(6)}[{\bf 50}] t^3 + \chi_{\so(6)}[{\bf 105}] t^4 + \ldots
\ee
where $t=x^{2-2{\bf r}}$ is the scaling dimension and the R-charge of the chiral ring generators. 
In terms of Dinkyn labels of $SU(4) \sim SO(6)$, we get an all-order expansion as sum over all representations with Dinkyn labels proportional to the Dinkyn label of the $\mathbf{6}$:%
\footnote{This is analogous to the Hilbert Series of the one-instanton moduli spaces \cite{Benvenuti:2010pq} which are sums over all irreps with Dinkyn labels proportional to the Dinkyn label of the adjoint representation.}
\be
\mathcal{HS} = \sum_{n=0}^{\infty} [0,n,0]_{su(4)}t^n \;.
\ee
Notice that only $SU(4)$ representations of even quadrality appear, consistent with the fact that the enhanced symmetry is $SO(6) = SU(4)/\bZ_2$.

\acknowledgments{We are grateful to Sara Pasquetti for very useful discussions. This work is supported in part by the MIUR-SIR grant RBSI1471GJ ``Quantum Field Theories at Strong Coupling: Exact Computations and Applications". S.B. is partly supported by the INFN Research Projects GAST and ST$\&$FI. }


\bibliographystyle{ytphys}

\end{document}